\documentclass[josab,osajnl,twocolumn,showpacs,superscriptaddress,10pt]{revtex4-1} 
\usepackage{amsmath,amssymb,graphicx}
\usepackage{palatino}
\usepackage{hyperref}

\begin{document}

\title{\rm  \bfseries Einstein--Podolsky--Rosen steering measure for two-mode continuous variable states}

\author{Ioannis Kogias}
\email{john$_$k$_$423@yahoo.gr}
\affiliation{School of Mathematical Sciences, The University of Nottingham,
University Park, Nottingham NG7 2RD, United Kingdom}
\affiliation{ICFO - The Institute of Photonic Sciences, Av. Carl Friedrich Gauss, 3 08860 Castelldefels (Barcelona), Spain}

\author{Gerardo Adesso}
\email{Gerardo.Adesso@nottingham.ac.uk}
\affiliation{School of Mathematical Sciences, The University of Nottingham,
University Park, Nottingham NG7 2RD, United Kingdom}

\begin{abstract}
Steering is a manifestation of quantum correlations that embodies the Einstein-Podolsky-Rosen (EPR) paradox. While there have been recent attempts to quantify steering, continuous variable systems remained elusive. We introduce a steering measure for two-mode continuous variable systems that is valid for arbitrary states. The measure is based on the violation of an optimized variance test for the EPR paradox by quadrature measurements, and admits a computable and experimentally friendly lower bound only depending on the second moments of the state, which reduces to a recently proposed quantifier of steerability by Gaussian measurements. We further show that Gaussian states are extremal with respect to our measure, minimizing it among all continuous variable states with fixed second moments.
As a byproduct of our analysis, we generalize and relate well-known EPR-steering criteria. Finally an operational interpretation is provided, as the proposed measure is shown to quantify a guaranteed key rate in semi-device independent quantum key distribution.
\end{abstract}

\maketitle 

\section{Introduction}
Almost 80 years have passed since the landmark paper of Einstein-Podolsky-Rosen (EPR) \cite{epr} on a paradoxical manifestation of quantum correlations which Schr\"odinger later termed {\it quantum steering} \cite{schr,schr2}, yet the topic is more timely than ever. From one-sided device independent entanglement verification \cite{wiseman} and quantum key distribution \cite{branciard,walk} to signifying secure quantum teleportation \cite{Reid13} and performing entanglement-assisted subchannel discrimination \cite{PianiSt}, Einstein's scrutinized notion of steering finds increasingly many applications in non-classical tasks after its recent formulation as a distinct type of asymmetric nonlocality by Wiseman and co-workers \cite{wiseman,wisepra}, thus making it a subject of intense research  \cite{eprpar}.

Steering, in a modern quantum information language \cite{wiseman,wisepra}, can be understood as the task of two distant parties, say Alice and Bob, in which Alice tries to convince Bob that the quantum state ${\hat{\rho}}_{AB}$ they share is entangled, by remotely creating quantum ensembles on Bob's site that could not have been created without shared entanglement. Given that Bob does not trust Alice and her announced measurements, we say that Alice can steer Bob's state (and thus convince Bob), or equivalently that the state ${\hat{\rho}}_{AB}$ is ``$A \to B$'' steerable, if and only if (\textit{iff}) the probabilities of all possible joint measurements cannot be written in the factorizable form \cite{wiseman}:
\begin{equation}\label{jointprob}
P \left( A,B|a,b,{\hat{\rho}}_{AB} \right) = \sum_{\lambda} {\cal P}_\lambda {\cal P}\left( A|a,\lambda \right) P\left(B|b,{\hat{\rho}}_\lambda \right),
\end{equation}
where the lower-case letters $a \in {\cal M}_A$ and $b \in {\cal M}_B$ denote local observables for Alice and Bob, while $A$ and $B$ their corresponding outcomes. Violation of \eqref{jointprob} implies the failure of a local hidden state model to explain the measurement statistics. As one can see from Eq.~\eqref{jointprob}, steering is an asymmetric form of nonlocality that sits in-between entanglement \cite{ent} and Bell nonlocality \cite{bell64,bell76,nonloc}. Not all entangled states are steerable, and not all steerable states are Bell nonlocal.

In order for steering to be useful one should first be able to detect it in experiments \cite{wittmann,saunders,eberle,handchen,bennet,smith,steinlechner,prydenew,seiji2015,icfo2015}. The first attempt to create an experimental criterion that captures the essence of the EPR paradox \cite{eprpar} in a continuous variable setting was made in the 80's by M. Reid \cite{reid}, whose criterion is commonly known as Reid's criterion and which was shown later to be only a special case of an EPR-steering test in the sense of \eqref{jointprob} \cite{cavalcanti}. Today our knowledge about the detection and distribution of steering  has significantly advanced \cite{asymmetry,he,reid13a}, with a plethora of effective criteria derived \cite{cavalcanti,walborn11,walborn13,cavalcanti2} and phenomena like steering monogamy identified in multi-party scenarios \cite{reid13a}. Besides  a yes/no answer to the question of steerability given by various steering criteria, however, one is interested in \textit{how much} a state is steerable for practical purposes.
Only quite recently, the quantification of steering was put forward by researchers \cite{Skrzypczyk,PianiSt,gausteer} to assess how much a quantum state's statistics deviate from \eqref{jointprob},  and thus how useful it can be for tasks that use steering as their resource \cite{resource}. Two measures of steering have been proposed in particular for finite-dimensional systems, the so-called steering weight \cite{Skrzypczyk}, and the steering robustness \cite{PianiSt}. While both measures are not amenable to analytical evaluation and can only be computed numerically by semidefinite programming, the steering robustness has a nice operational interpretation in the context of subchannel discrimination \cite{PianiSt}. For continuous variable systems, a computable steering quantifier specific to Gaussian states and measurements has also been very recently proposed \cite{gausteer}.

In this paper we present an accessible approach to the quantitative estimation of steerability for bipartite two-mode continuous variable states. We examine recent experimental criteria for steering \cite{cavalcanti}, the so-called EPR-Reid variance criteria whose applicability extends to all (Gaussian and non-Gaussian) states, and analyze their maximal violation by optimal local quadrature observables for Alice and Bob, in order to capture the largest possible departure from  \eqref{jointprob} for a given state. Hence we define (in Section 2) a suitable measure of steering for an arbitrary two-mode state, and we prove that it admits an analytically computable lower bound that captures the degree of steerability of the given state by Gaussian measurements. The lower bound coincides with the Gaussian steering measure introduced in a previous work \cite{gausteer}, whose usefulness is here generalized from the Gaussian domain to arbitrary states.
We prove Gaussian states to be in fact extremal \cite{extremality}, as they are minimally steerable among all states with the same covariance matrix, according to the  measure proposed in this paper.
As a corollary of our analysis, we show (in Section 3) that a necessary and sufficient condition for steerability of Gaussian states under Gaussian measurements obtained by Wiseman \textit{et al.} based on covariance matrices \cite{wiseman,wisepra}, remains valid  as a sufficient steering criterion for arbitrary non-Gaussian states,  and amounts to Reid's criterion \cite{reid,eprpar} when optimal Gaussian local observables are chosen for the latter. We conclude (in Section 4) with a summary of our results and an outlook of currently open questions motivated by the present analysis.

\section{A steering measure for two-mode states based on quadrature measurements}
In general \cite{bennett}, a measure of steering should quantify how much the correlations of a quantum state depart from the expression in Eq.~\eqref{jointprob}. Since a manifestation of these correlations can be observed by the violation of suitable EPR-steering criteria, one can get a quantitative estimation of the degree of steerability in a given state by evaluating the maximum violation of a chosen steering criterion as revealed by  optimal measurements. One expects that the higher the violation (i.e., the amount of correlations), the more useful the state will be in tasks that use quantum steering as a resource.

In this paper we consider an arbitrary state ${\hat{\rho}}_{AB}$ of a two-mode continuous variable system.
The relevant steering criteria to our work will be the so-called multiplicative variance EPR-steering criteria \cite{cavalcanti}, of which  Reid's criterion \cite{reid} is a special case. Following \cite{cavalcanti,reid,eprpar}, let us consider a situation where Bob measures two canonically conjugate observables on his subsystem, $\hat{x}_B,
\hat{p}_B$ with corresponding outcomes $X_B, P_B$, and Alice tries to guess Bob's outcomes based on the  outcomes 
of measurements 
on her own subsystem.
If, say, the outcome of Alice's measurement is $X_A$, corresponding to a local observable $\hat{x}_A$,
we can denote by ${X_{\rm est}}\left( X_A \right)$ Alice's inference of Bob's measurement outcome $X_B$. The average inference variance of $X_B$ given Alice's estimator  ${X_{\rm est}}\left( X_A \right)$ is defined by
\begin{equation}\label{infvar}
{\Delta ^2_{\inf}}X_B = \left\langle {{{\left[ {X_B - {X_{\rm{\rm est}}}\left( X_A \right)} \right]}^2}} \right\rangle,
\end{equation}
where the average is taken with respect to the joint probability distribution $P \left(X_A, X_B\right)$ and over all outcomes $X_A,X_B$. One can show \cite{eprpar} that the optimal estimator minimizing the inference variance $\Delta^2_{\rm inf}X_B$ is the mean $X_{\rm est}\left(X_A\right)=\left\langle X_B \right\rangle_{X_A}$ evaluated on the conditional distribution $P\left(X_B|X_A \right)$. Substituting in \eqref{infvar} we obtain the minimal inference variance of $X_B$ by measurements on $A$,
\begin{equation}\label{optinfvar}
\Delta^2_{\min}X_B = \sum\nolimits_{X_A} {P\left( X_A \right){\Delta ^2}} \left( {X_B|X_A} \right)\,,
\end{equation}
where $\Delta^2 \left(X_B|X_A\right)$ is the conditional variance of $X_B$ calculated from $P\left(X_B |X_A \right)$. Clearly, from the properties stated above, it holds that
$
\Delta^2_{\inf}X_B \geq \Delta^2_{\min}X_B.$
Similarly we can define an inference variance $\Delta^2_{\inf}P_B$ for $\hat{p}_B$ and its corresponding minimum $\Delta^2_{\min}P_B$  given respectively by analogous formulas to \eqref{infvar} and \eqref{optinfvar}, but conditioned on $P_A$ instead of $X_A$. In \cite{cavalcanti,eprpar} it was shown that a bipartite state ${\hat{\rho}}_{AB}$ shared by Alice and Bob is steerable by Alice, i.e. ``$A \to B$'' steerable, if the condition
\begin{equation}\label{EPRcrit}
\Delta^2_{\min}X_B \,\, \Delta^2_{\min}P_B \geq 1,
\end{equation}
is violated.

 Notice that the criterion \eqref{EPRcrit} is independent of Alice's and Bob's first moments, since displacements of the form $X_{A(B)} \to X_{A(B)} + d_{A(B)}$ leave the inference variances (of both position and momentum) invariant as can be easily seen from the definition \eqref{optinfvar}. Therefore, first moments will be assumed to be zero in the rest of the paper without any loss of generality.

We remark that the EPR-steering criterion \eqref{EPRcrit} is applicable to arbitrary states and is valid without any assumption on the Hilbert space of Alice's subsystem, as Bob just needs to identify two distinctly labelled measurements performed by Alice \cite{cavalcanti}. However, in order to keep our analysis accessible, we will further assume that Alice's allowed measurements are restricted to be quadrature ones, i.e., projections on the eigenbasis of (generally rotated) canonically conjugate operators $\hat{x}^\theta_A$ and $\hat{p}^\theta_A$, such that $[\hat{x}^\theta_A, \hat{p}^\theta_A]=i$ in natural units. Although quadrature measurements are not general and not necessarily optimal to detect steerability in all states, they are convenient from a theoretical point of view and can be reliably implemented in laboratory by means of homodyne detections.

One immediately sees that the product of variances in \eqref{EPRcrit} is not invariant under local unitary operations (apart from displacements) by Alice and Bob, thus a state might be detected as more or less steerable if some local change of basis is implemented.  In order to capture steerability in an invariant way, one can consider the maximum violation of \eqref{EPRcrit} that a quantum state  $\hat{\rho}_{AB}$ can exhibit, by minimizing the product $\Delta^2_{\min}X_B \,\, \Delta^2_{\min}P_B$ over all local unitaries $U_{\rm{local}}=U_A \otimes U_B$ for $A$ and $B$ applied to the state.

We then propose to \textit{quantify} the ``$A \to B$'' steerability of an arbitrary two-mode continuous variable state ${\hat{\rho}}_{AB}$ detectable by quadrature measurements, via the measure
\begin{equation}\label{measure}
{\cal S} ^{A \to B}\left({\hat{\rho}}_{AB}\right) = \max \left\lbrace 0,\,-\frac12
\ln {\cal F} \right\rbrace,
\end{equation}
where \begin{equation}\label{fuffa}
{\cal F} = \min_{\lbrace U_{\rm{local}} \rbrace} \Delta^2_{\min}X_B \, \Delta^2_{\min}P_B.
\end{equation}
The measure naturally quantifies the amount of violation of an optimized multiplicative variance EPR-steering criterion of the form \eqref{EPRcrit} for an arbitrary state ${\hat{\rho}}_{AB}$. As one would expect from any proper quantifier of quantum correlations, the measure enjoys local unitary invariance by definition, and it vanishes for all states which are not ``$A \to B$'' steerable.

Calculating ${\cal S} ^{A \to B}$  in an analytical manner for an arbitrary state is still a difficult task. In general, given a quantum state, the minimization in $\cal F$ involves both Gaussian and non-Gaussian local unitaries for Alice and Bob, which correspond to violations of (\ref{EPRcrit}) by Gaussian and non-Gaussian quadrature measurements, respectively.
It is possible, though, to obtain a computable lower bound to ${\cal S} ^{A \to B}$  if one constrains the optimization to Gaussian unitaries only.
The lower bound, presented in the next subsection, will then provide a quantitative indication of the ``$A \to B$'' steerability of ${\hat{\rho}}_{AB}$ that can be demonstrated by Gaussian  measurements on Alice's subsystem.

\subsection{Lower bound}

A short introduction of the reader to Gaussian states is first intended \cite{ournewreview}. An arbitrary bipartite Gaussian state ${\hat{\rho}^G}_{AB}$ is  determined, up to local displacements, by its second moments, i.e., it is specified  by the covariance matrix (CM) $\sigma_{AB}$, which can be written in the block form
\begin{equation}
\label{coma}
\sigma_{AB}=\left(\begin{matrix}
A & C \\
C^T & B \\
\end{matrix}\right).
\end{equation}
Here,  $A$ and $B$ are the marginal CMs corresponding to the reduced states of Alice and Bob respectively, while $C$ encodes intermodal correlations. For two-mode states, $A$, $B$, and $C$ are $2 \times 2$ matrices.
The matrix elements of the CM, defined by $(\sigma_{AB})_{ij}={\rm Tr}\big[ \left( \hat{R}_i \hat{R}_j + \hat{R}_j \hat{R}_i\right) {\hat{\rho}^G}_{AB}\big]$, are expressed via the vector $\hat{R}= \left( \hat{x}_A,\hat{p}_A,\hat{x}_B, \hat{p}_B \right)^T$ that conveniently groups the phase-space operators $\hat{x}_{A(B)},\hat{p}_{A(B)}$ for each mode. The canonical commutation relations these operators satisfy  can be compactly expressed as  $[\hat{R}_j,\hat{R}_k]=i(\Omega_{AB})_{jk}$, where $\Omega_{AB}= \Omega_{A} \oplus \Omega_{B}$ is the symplectic matrix, with $\Omega_{A} = \Omega_{B} = \left( \begin{array}{*{20}{c}}
0 & 1 \\
-1 & 0 \\
\end{array}\right)$  \cite{ournewreview}. The CM of any (Gaussian or non-Gaussian) physical state needs to satisfy the \textit{bona fide} condition
\begin{equation}\label{bonafide}
\sigma_{AB} + i \left(\Omega_A \oplus \Omega_B \right) \geq 0\,.
 \end{equation}
Gaussian operations are defined as those which preserve the Gaussianity of the states they act upon.

To obtain a lower bound for the steering measure ${\cal S} ^{A \to B}\left({\hat{\rho}}_{AB}\right)$ in terms of second moments, we first remind the reader that with no loss of generality one can assume vanishing first moments of ${\hat{\rho}}_{AB}$ (see discussion below Eq. \eqref{EPRcrit}). We will show that, for arbitrary states ${\hat{\rho}}_{AB}$ with corresponding CM $\sigma_{AB}$, the product of inference variances $\Delta^2_{\inf} X_B\, \Delta^2_{\inf} P_B$, defined as in (\ref{infvar}), acquires its minimum value when $\sigma_{AB}$ is expressed in the so-called standard form
\begin{equation}\label{stanform}
{\bar{\sigma} _{AB}} = \left( \begin{matrix}
\bar{A} & \bar{C} \\
\bar{C}^T & \bar{B} \\
\end{matrix}\right),
\end{equation}
in which the submatrices $\bar{A}={\rm diag}\left( a,a \right)$, $\bar{B}={\rm diag}\left(b,b\right)$, and $\bar{C}={\rm diag}\left(c_1,c_2\right)$ take a diagonal form. The standard form can always be obtained for any state by suitable local unitary operations \cite{simon00,duan00} and is unique up to a sign flip in $c_1$ and $c_2$, as its elements can be recast as functions of four local invariants of the CM \cite{extremal}.

Let us begin by considering a steerable ${\hat{\rho}}_{AB}$ that violates \eqref{EPRcrit}, so that ${\cal S}^{A \to B}\left({\hat{\rho}}_{AB}\right)>0$. We use the fact that $\Delta^2_{\inf}X_B \geq \Delta^2_{\min}X_B$, when a linear estimator $X_{\rm{\rm est}} \left( X_A \right)=g_x X_A +d_x$ is used in \eqref{infvar}; after minimizing the inference variance over the real numbers $g_x, d_x$ and considering vanishing first moments without any loss of generality, we find $\Delta^2_{\inf}X_B = \langle X_B^2 \rangle -\langle X_B X_A \rangle^2 /\langle X_A^2 \rangle$ \cite{eprpar}. Similar considerations hold for the inference variance of momentum, where an estimator of the form $P_{\rm{\rm est}}\left(P_A\right)=g_p P_A+d_p$ will give
$\Delta^2_{\inf}P_B = \langle P_B^2 \rangle -\langle P_B P_A \rangle^2 /\langle P_A^2 \rangle$ after optimizing over the real numbers $g_p,d_p$.

Since a linear estimator is optimal for inferring the variance in the case of Gaussian states \cite{reid,eprpar}, but not anymore in the general case, the inequality  $\Delta^2_{\inf}X_B\Delta^2_{\inf}P_B \geq \Delta^2_{\min}X_B \Delta^2_{\min}P_B$ will be true for all states (with equality on Gaussian states). Hence, $\cal F$ in \eqref{measure} can be upper bounded as follows,
\begin{equation}\label{F}
\begin{split}
{\cal F} & =\min_{\lbrace U_{G}\rbrace \cup \lbrace U_{nG} \rbrace}\Delta^2_{\min}X_B \Delta^2_{\min}P_B \\
& \leq \min_{\lbrace U_{G}\rbrace \cup \lbrace U_{nG} \rbrace}\Delta^2_{\inf}X_B \Delta^2_{\inf}P_B \\
& \leq \min_{\lbrace U_{\rm{G}}\rbrace}\, \Delta^2_{\inf}X_B \Delta^2_{\inf}P_B,
\end{split}
\end{equation}
where we have decomposed the set of local unitaries $\lbrace U_{\rm{local}}\rbrace$ into Gaussian $\lbrace U_G \rbrace$ and non-Gaussian $\lbrace U_{nG} \rbrace$ ones.
%
The product of inference variances in \eqref{F}  is intended as evaluated from the optimal linear estimator as detailed above \cite{eprpar}, namely
\begin{equation}\label{Reid}
\begin{split}
 \Delta^2_{\inf}X_B  \Delta^2_{\inf}P_B & = \left(\langle X_B^2 \rangle -\langle X_B X_A \rangle^2 /\langle X_A^2 \rangle \right) \times \\
&  \left(  \langle P_B^2 \rangle -\langle P_B P_A \rangle^2 /\langle P_A^2 \rangle \right),
\end{split}
\end{equation}
Since an upper bound on ${\cal F}$ will give us the desired lower bound on ${\cal S}^{A \to B}$, what remains is to compute this upper bound, i.e., the rightmost quantity in (\ref{F}), which only depends on the CM elements of the state. Note that the product of inference variances \eqref{Reid}, using linear estimators, defines what is well-known in the literature as Reid's criterion \cite{reid},
\begin{equation}\label{ReReid}
\Delta^2_{\inf}X_B  \Delta^2_{\inf}P_B \geq  1\,,
\end{equation}
whose violation is sufficient to detect ``$A \to B$'' steerability of a general two-mode state based on second order moments.

Local Gaussian unitaries (that do not give rise to displacements) acting on states $\hat{\rho}_{AB}$, translate on the level of CMs as local symplectic transformations $S_{\rm{local}}=S_A \oplus S_B$, acting by congruence: $\sigma_{AB} \mapsto S_{\rm{local}} \sigma_{AB} S_{\rm{local}}^T$ \cite{simon94,ournewreview}.
In order to compute $\min_{\lbrace S_{\rm{local}} \rbrace}\Delta^2_{\inf}X_B \Delta^2_{\inf}P_B$ we can, with no loss of generality, consider a CM $\bar{\sigma}_{AB}$ in standard form, apply an arbitrary local symplectic operation $S_{\rm{local}}$ to it, then evaluate $\Delta^2_{\inf}X_B\, \Delta^2_{\inf}P_B$ on the transformed CM $S_{\rm{local}}\bar{\sigma}_{AB}S_{\rm{local}}^T$, and finally minimize this quantity over all possible matrices $S_{A(B)}$. To perform the minimization we parametrize the matrix elements of $S_{A(B)}$ in the following convenient way,
\begin{equation}\label{parametr}
S_{A(B)}  = \left( {\begin{array}{*{20}{c}}
   {\frac{1}{{\left( {1 - {u_{A(B)}}{v_{A(B)}}} \right)w_{A(B)}}}} & {\frac{{{v_{A(B)}}}}{{\left( {1 - {u_{A(B)}}{v_{A(B)}}} \right)w_{A(B)}}}}  \\
   {{u_{A(B)}}w_{A(B)}} & {w_{A(B)}}  \\
\end{array}} \right)
\end{equation}
where the symplectic condition $S_{A(B)} \Omega_{A(B)} S^T_{A(B)} = \Omega_{A(B)}$  has been taken into account and the real variables $u_{A(B)}, v_{A(B)}, w_{A(B)}$ are now independent of each other. Performing the (unconstrained) minimization over the variables $u_{A(B)},v_{A(B)}$ we were able to obtain analytically the global minimum of the product (\ref{Reid}) with respect to Gaussian observables,
\begin{equation}\label{globalmin}
\min_{\lbrace U_G \rbrace}\left[\Delta^2_{\inf}X_B\, \Delta^2_{\inf}P_B \right] = \det M^B_{\sigma},
\end{equation}
which also constitutes the upper bound for $\cal F$ in \eqref{F}. Here the local symplectic invariant $\det M^B_{\sigma}= \left( b-\frac{c_1^2}{a}\right)\left( b-\frac{c_2^2}{a}\right)$ is the determinant of the Schur complement of $A$ in ${\sigma}_{AB}$, defined for any two-mode CM (\ref{coma}) as \cite{wisepra,gausteer}
\begin{equation}\label{mubs}
M_\sigma ^B = B - {C^T}{A^{ - 1}}C\,.
\end{equation}
The minimum \eqref{globalmin} can be obtained from every state using the following parameters that determine the local symplectic operations \eqref{parametr},
$$\left( {{u_A},{v_A},{u_B},{v_B}} \right) = \left( {\begin{array}{*{10}{c}}
   {\frac{{{c_1}{v_B}}}{{{c_2}}},} \!&\! {\frac{{ - ab + c_1^2}}{{ab - c_2^2}}\frac{{{c_2}{v_B}}}{{{c_1}}},} \!&\! {\frac{{ - ab + c_1^2}}{{ab - c_2^2}}{v_B},} \!&\! {{v_B}}  \\
\end{array}} \right),$$
$\forall$ $v_B, w_{A(B)}$. It is evident from \eqref{globalmin} that the minimum product of inference variances \eqref{Reid} is achieved, in particular, when evaluated for a standard form CM $\bar{\sigma}_{AB}$.

Substituting
${\cal F} \leq \det M^B_\sigma$ in \eqref{measure}, a lower bound for the proposed steering measure of an arbitrary two-mode state $\hat{\rho}_{AB}$ is obtained,
\begin{equation}\label{lowerbound}
{\cal S}^{A \to B}\left({\hat{\rho}}_{AB}\right) \geq {\cal G}^{A \to B}\left( \sigma_{AB} \right),
\end{equation}
where we recognize the Gaussian steering measure introduced in \cite{gausteer},
\begin{equation}\label{gausteer}
{\cal G}^{A \to B}\left( \sigma_{AB} \right)=\max \left\lbrace 0,  -
\frac12\ln \det M_\sigma ^B \right\rbrace.
\end{equation}
The lower bound ${\cal G}^{A \to B}$ solely depends on local symplectic invariant quantities that uniquely specify the CM of the state. As is known \cite{extremal}, these invariant quantities can be expressed back with respect to the original elements of the CM which one can measure in laboratory, e.g.~via homodyne tomography \cite{francamentemeneinfischio}. Henceforth, the lower bound that we obtained is both analytically computable and, also, experimentally accessible in a routinely fashion for any (Gaussian or non-Gaussian) state, since only moments up to second order are involved.

In the following we discuss some useful properties that the steering measure ${\cal S}^{A \to B}$ and its lower bound ${\cal G}^{A \to B}$ satisfy, and show how these results can be used to link and generalize  existing steering criteria.

\subsection{Properties}

In a recent work \cite{gausteer} the present authors introduced a measure  of EPR-steering for multi-mode bipartite Gaussian states that dealt with the problem of ``\textit{how much a Gaussian state can be steered by Gaussian measurements}". This measure ${\cal G}^{A \to B}$ was defined as the amount of violation of the following criterion by Wiseman \textit{et al.} \cite{wiseman,wisepra},
\begin{equation}\label{nonsteer}
{\sigma _{AB}} + i\,({0_A} \oplus {\Omega _B}) \ge 0.
\end{equation}
Violation of (\ref{nonsteer}) gives a necessary and sufficient condition for ``$A \to B$'' steerability of Gaussian states by Gaussian measurements. We recall from the original papers \cite{wiseman, wisepra}, where the details can be found, that for two modes the condition \eqref{nonsteer} is violated \textit{iff} $\det M_\sigma^B<1$, hence equivalently \textit{iff} ${\cal G}^{A \to B}\left( \sigma_{AB} \right) > 0$, where the Gaussian steering measure is defined in (\ref{gausteer}).
In a two-mode continuous variable system, a non-zero value of Gaussian steering ${\cal G}^{A \to B}>0$ detected on a CM $\sigma_{AB}$, which implies a non-zero value of the more general measure ${\cal S}^{A \to B}>0$ due to \eqref{lowerbound}, constitutes therefore not only a necessary and sufficient condition for the steerability by Gaussian measurements of the Gaussian state $\hat{\rho}_{AB}^G$ defined by $\sigma_{AB}$, but also a sufficient condition for the steerability of all (non-Gaussian) states $\hat{\rho}_{AB}$ with the same CM $\sigma_{AB}$.


While ${\cal S}^{A\to B}$ is hard to study in complete generality, its lower bound however has been shown to satisfy a plethora of valuable properties. In \cite{gausteer} we showed that Gaussian steering acquires for two modes a form of coherent information \cite{wilde}, ${\cal G}^{A\to B}\left(\sigma_{AB}\right) = \max \lbrace 0, {\cal S}\left(A \right)- {\cal S} \left( \sigma_{AB} \right) \rbrace$, with the R\'enyi-2 entropies ${\cal S}\left( \sigma \right) = \frac{1}{2}\ln \left( \det \sigma \right)$ \cite{renyi} replacing the standard von Neumann ones. Thanks to this connection ${\cal G}^{A\to B}\left(\sigma_{AB}\right)$ was shown to satisfy various  properties that we repeat here without proof: (a) ${\cal G}^{A \to B}\left( \sigma_{AB} \right)$ is convex and additive; (b) ${\cal G}^{A \to B}\left( \sigma_{AB} \right)$  is monotonically decreasing under Gaussian quantum operations on the (untrusted) party Alice; (c) ${\cal G}^{A \to B}\left( \sigma_{AB} \right) = {\cal E}\left(\sigma^p_{AB}\right)$ for $\sigma^p_{AB}$ pure, and, (d) ${\cal G}^{A \to B}\left( \sigma_{AB} \right) \leq {\cal E}\left(\sigma_{AB}\right)$ for $\sigma_{AB}$ mixed, where ${\cal E}$ denotes the Gaussian Renyi-2 entropy measure of entanglement \cite{renyi}. In the light of the recently developed resource theory of steering \cite{resource} properties (a) and (b) should be satisfied by any proper measure of steering, while properties (c) and (d) should be satisfied by any quantifier that respects the hierarchy of quantum correlations.
The present paper, thus, validates all the already established properties of ${\cal G}^{A \to B}$ as an indicator of steerability by Gaussian measurements, and extends them to arbitrary states.

Interestingly, (\ref{lowerbound}) suggests that by accessing only the second moments of an arbitrary state, one will not overestimate its steerability according to our measure. We can make this observation rigorous by showing that the steering quantifier ${\cal S}^{A \to B}$ satisfies an important {\it extremality} property as formalized in \cite{extremality}. Namely, the Gaussian state $\hat{\rho}^G_{AB}$ defined by its CM $\sigma_{AB}$ minimizes ${\cal S}^{A \to B}$ among all states $\hat{\rho}_{AB}$ with the same CM $\sigma_{AB}$. This follows by recalling that the Reid product (\ref{Reid}), which appears in (\ref{F}), is independent from the (Gaussian versus non-Gaussian) nature of the state, and that linear inference estimators are globally optimal for Gaussian states as mentioned above \cite{eprpar}. This entails that the middle term in (\ref{F}) can be recast as
\begin{equation}\label{FG}
\begin{split}
& \min_{\lbrace U_{G}\rbrace \cup \lbrace U_{nG} \rbrace}(\Delta^2_{\inf}X_B \Delta^2_{\inf}P_B)_{\hat{\rho}_{AB}} \\
= & \min_{\lbrace U_{G}\rbrace \cup \lbrace U_{nG} \rbrace}(\Delta^2_{\inf}X_B \Delta^2_{\inf}P_B)_{\hat{\rho}_{AB}^G} \\
= & \min_{\lbrace U_{G}\rbrace \cup \lbrace U_{nG} \rbrace}(\Delta^2_{\min}X_B \Delta^2_{\min}P_B)_{\hat{\rho}_{AB}^G} \\
= &\ \  {\cal F}(\hat{\rho}^G_{AB})\,,
\end{split}
\end{equation}
where, for the sake of clarity, we have explicitly indicated the states on which the variances are calculated: $\hat{\rho}_{AB}$ denotes an arbitrary two-mode state, and $\hat{\rho}_{AB}^G$ corresponds to the reference Gaussian state with the same CM.

Therefore, combining Eqs.~(\ref{measure}), (\ref{F}), (\ref{lowerbound}), and (\ref{FG}), we can write the following chain of inequalities for the ``$A \to B$'' steerability of an arbitrary two-mode state $\hat{\rho}_{AB}$,
\begin{equation}\label{extremalityistrue}
{\cal S}^{A \to B}\left({\hat{\rho}}_{AB}\right) \geq {\cal S}^{A \to B}\left({\hat{\rho}^G}_{AB}\right) \geq {\cal G}^{A \to B}\left( \sigma_{AB} \right)\,.
\end{equation}
The leftmost inequality in (\ref{extremalityistrue}) embodies the desired extremality property \cite{extremality} for our steering measure.
This is very relevant in a typical experimental situation, where the exact nature of the state $\hat{\rho}_{AB}$ is mostly unknown to the experimentalist. Then, thanks to (\ref{extremalityistrue}) we rest assured that, by assuming a Gaussian nature of the state under scrutiny, the experimentalist will never overestimate the EPR-steering correlations between Alice and Bob as quantified by the measure defined in (\ref{measure}).

Finally, coming to operational interpretations for our proposed steering quantifier ${\cal S}^{A \to B}$, we show that it is connected to the figure of merit of semi-device independent quantum key distribution \cite{walk}, that is, the secret key rate.  In the conventional entanglement-based quantum cryptography protocol \cite{ekert}, Alice and Bob share an arbitrary two-mode state ${\hat{\rho}}_{AB}$, and want to establish a secret key given that Alice does not trust her devices. By performing local measurements (typically homodyne detections) on their modes, and a direct reconciliation scheme (where Bob sends corrections to Alice)  they can achieve the secret key rate \cite{walk}
\begin{equation}
\label{kwk}
K \geq \max \left\lbrace 0, \ln \left( \frac{2}{e \sqrt{\Delta^2_{\inf}X_B \Delta^2_{\inf}P_B}} \right) \right\rbrace
\,.
\end{equation}
Notice that the secret key rate depends on the expression in (\ref{Reid}), which is not unitarily invariant. Therefore, it can be optimized over local unitary operations. In the case where $\Delta^2_{\inf} X_B \Delta^2_{\inf} P_B$ takes its minimum value for the given shared ${\hat{\rho}}_{AB}$, the lower bound on the correspondingly optimal key rate $K_{\rm opt}$ can be readily expressed in terms of the ``$A \to B$'' steerability measure, yielding
\begin{equation}\label{keyrate}
K_{\rm opt} \geq \max \big\lbrace 0,\, {\cal S}^{A \to B}\left({\hat{\rho}}_{AB}\right)+\ln 2-1  \big\rbrace.
\end{equation}
Thus, ${\cal S}^{A \to B}$ quantifies a guaranteed key rate for any given state. If a reverse reconciliation protocol is used (in which Alice sends corrections to Bob) the quantifier ${\cal S}^{B \to A}$ of the inverse steering direction enters \eqref{keyrate} instead. Thus, one sees that the asymmetric nature of steering correlations can play a decisive role in communication protocols that rely on them as resources. In the cryptographic scenario discussed, if the shared state ${\hat{\rho}}_{AB}$ is only one-way steerable, say $A \to B$, then a reverse reconciliation protocol that relies on ${\cal S}^{B \to A}$ is not possible. A looser lower bound to the key rate \eqref{keyrate} can also be expressed in terms of ${\cal G}^{A \to B}$ by using \eqref{lowerbound}, in case one wants to study the advantage that Gaussian steering alone gives for the key distribution, or one just wants to get an estimate.

\begin{figure}[t!]
\includegraphics[width=\columnwidth]{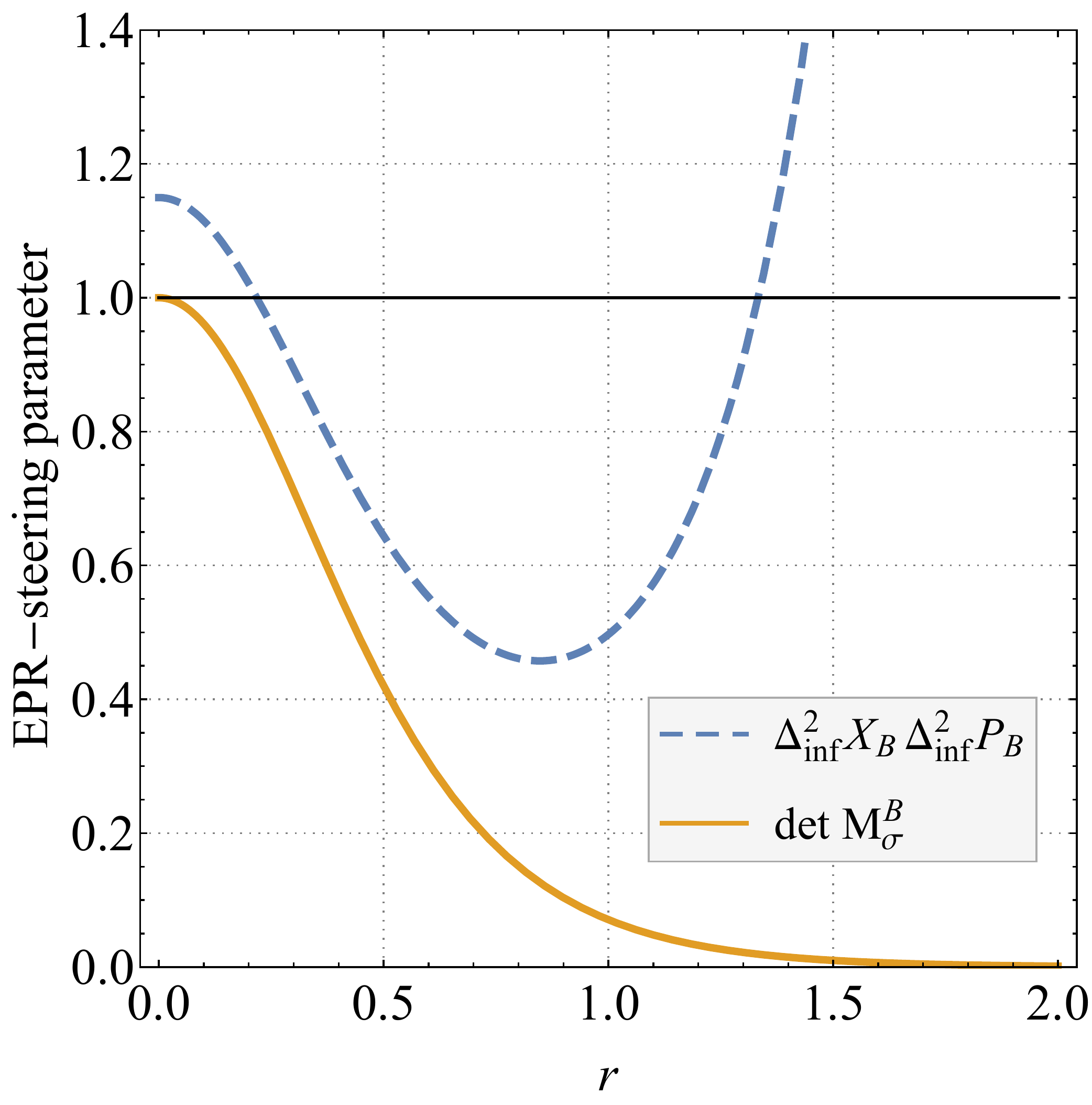}
\caption{(Color online)
We illustrate the performance of Reid's \cite{reid} and Wiseman \textit{et al.}'s \cite{wiseman} EPR-steering criteria for the steering detection of a pure two-mode squeezed state with squeezing $r$, with CM transformed from the standard form by the application of a local symplectic transformation parameterized as in (\ref{parametr}), with $u_{A(B)} = v_{A(B)}/(1+v_{A(B)}^2)$, $w_{A(B)} = 1 + v_{A(B)}^2$ (in the plot, we choose $v_{A}=0.16$ and $v_{B}=0.19$). The criteria are represented by their figures of merit, namely the product of conditional variances (dashed blue line) for Reid's criterion (\ref{ReReid}) and the determinant $\det M_B$ (solid orange line) for  Wiseman \textit{et al.}'s criterion (\ref{nonsteer}). The two-mode squeezed state is steerable for all $r>0$, but the aforementioned criteria detect this steerability only when their respective parameters give a value smaller than unity (straight black line). As one can see, we have $\det M_B<1$ for all $r>0$ and independently of any local rotations, while Reid's criterion detects steerability only for a small range of squeezing degrees and is highly affected by local rotations. If the state is sufficiently rotated out of the standard form, the unoptimized Reid's criterion will not be able to detect any steering at all.}
\label{wisereid}
\end{figure}

\section{Reid, Wiseman, and a stronger steering criterion}

Finally, we discuss the implications of our work on existing EPR-steering criteria \cite{reid,wiseman}.
The second order EPR-steering criteria by Reid \eqref{ReReid} and Wiseman \textit{et al.} \eqref{nonsteer}, are perhaps the most well-known ones for continuous variable systems. Although a comparison between them has been issued before in a special case (two-mode Gaussian states in standard form) \cite{wisepra}, they appear to exhibit quite distinct features in general \cite{cavalcanti}. On one hand, Wiseman \textit{et al.}'s criterion (\ref{nonsteer}), defined only in the Gaussian domain, is invariant under local symplectics and provides a necessary and sufficient condition for steerability of Gaussian states under Gaussian measurements. On the other hand, Reid's criterion (\ref{ReReid}) is applicable to all states but is not invariant under local symplectics and as a result it cannot always detect steerability even on a Gaussian state. As an illustrative example, we show in Fig.~\ref{wisereid} the performance of the two criteria for steering detection in a pure two-mode squeezed state, locally rotated out of its standard form. One can clearly see that Wiseman \textit{et al.}'s criterion is superior to the non-optimized Reid's one, which fails to detect steering in the regimes of very low or very high squeezing \cite{Note1}.

However, it was previously argued \cite{cavalcanti} that Wiseman {\it et al.}'s stronger condition could not qualify as a general steering criterion, and could not be used in an experimental scenario where sources of non-Gaussianity may be present, since the derivation of the criterion and its validity were limited strictly to the Gaussian domain, while general EPR-steering criteria should be defined for all states and measurements. The exact connection established by \eqref{globalmin} between Wiseman \textit{et al.}'s figure of merit, $\det M^B_\sigma$, and Reid's product of inference variances (\ref{Reid}), makes us realize now that the two criteria are just two sides of the same coin; i.e., Wiseman \textit{et al.}'s  criterion represents the best performance of Reid's criterion when optimal Gaussian observables are used for the latter. As a byproduct of this connection,  we have thus upgraded the validity of Wiseman \textit{et al.}'s criterion to arbitrary two-mode continuous variable states. Namely, our results imply that  a violation of (\ref{nonsteer}) on any state $\hat{\rho}_{AB}$ with CM $\sigma_{AB}$ is sufficient to certify its ``$A \to B$'' steerability, as detectable in laboratory by optimal quadrature measurements. This condition can be thus regarded, to the best of our current knowledge, as the strongest experimentally friendly EPR-steering criterion for arbitrary two-mode states involving moments up to second order.

\section{Conclusion}

In conclusion, we introduced a quantifier of EPR-steering for arbitrary bipartite two-mode continuous-variable states, that can be estimated both experimentally and theoretically in an analytical manner. Gaussian states were found to be extremal with respect to our measure, minimizing it among all continuous variable states with fixed second moments \citep{extremality}. By further restricting to Gaussian measurements, we obtained a computable lower bound for any (Gaussian or non-Gaussian) two-mode state, that was shown to satisfy a plethora of good properties \cite{gausteer}. The measure proposed in this paper is seen to naturally quantify a guaranteed key rate for semi-device independent quantum key distribution \cite{walk}. Finally, this work generalizes and sheds new light on existing steering criteria based on quadrature measurements \cite{reid,wiseman}.

Nevertheless many questions still remain, complementing the ones posed previously in \cite{gausteer}. To begin with, it would be worthwhile to extend the results presented here to multi-mode states and see whether a connection similar to \eqref{globalmin} still holds. We also leave for further research the possibility that our quantifier (or its lower bound) may enter in other figures of merit for protocols that consume steering as a resource, like the tasks of secure quantum teleportation and teleamplification of Gaussian states \cite{Reid13,reid14} or entanglement-assisted Gaussian subchannel discrimination with one-way measurements \cite{PianiSt}.
Moreover, the proved connection of the measure with entropic quantities in the purely Gaussian scenario could be an instance of a more general property that we believe is worth investigating, possibly making the link with the degree of violation of more powerful (nonlinear) entropic steering tests \cite{walborn11,walborn13}.

Finally,  it is presently unknown whether the rightmost inequality in (\ref{extremalityistrue}) is tight; namely, whether or not non-Gaussian unitaries in the minimization of (\ref{measure}) can give rise to higher steerability of Gaussian states, compared to optimal Gaussian unitaries. This is related to the open question, first posed in \cite{wiseman}, of whether or not there exist steerable Gaussian states which nonetheless cannot be steered by Gaussian measurements; so far, such states have not been found even by resorting to nonlinear steering criteria \cite{walborn11,kimnha}.  On one hand, one would expect that Gaussian measurements are optimal for steering Gaussian states, since Gaussian operations and decompositions are indeed optimal for (provably a large class of) two-mode Gaussian states when entanglement and discord-type correlations are considered \cite{gaussianeof,giovaentanglement,gaussiandiscord,pirandoladiscord}. On the other hand, non-Gaussian measurements are always required to violate any Bell inequality on Gaussian states \cite{nhaviol, cerfviol} by virtue of their positive Wigner function, hence Gaussian measurements are in contrast completely useless for that task. Since steering is the `missing link' which sits just below nonlocality and just above entanglement in the hierarchy of quantum correlations \cite{wiseman,wisepra}, pinning down precisely the role of Gaussian measurements for steerability of Gaussian states would be particularly desirable.
Here, we dare to conjecture that ${\cal S}^{A \to B}(\hat{\rho}_{AB}^G) = {\cal G}^{A \to B}(\sigma_{AB})$, that is, that the general measure of EPR-steering introduced in this paper would reduce exactly to the measure of Gaussian steering proposed in \cite{gausteer}, for all two-mode Gaussian states; this would signify the optimality of Gaussian measurements for steerability of Gaussian states. However, a proof or disproof of this tempting hypothesis is beyond our current capabilities, and is left here as a future challenge to the community.

\acknowledgements{
This work was supported by the University of Nottingham (International Collaboration Fund) and the Foundational Questions Institute (FQXi-RFP3-1317). We are grateful to A. Ac\'in, D. Cavalcanti, E. G. Cavalcanti, A. Doherty, Q. Y. He, A. R. Lee, M. Piani, S. Ragy, M. D. Reid, P. Skrzypczyk, and especially H. Wiseman for valuable discussions.}


\bibliographystyle{osajnl}
\bibliography{SteeringnewBib}

\end{document}